\documentclass[a4paper]{article}
\usepackage{amsmath,amsfonts}
\usepackage[dvips]{graphicx}
\usepackage[bookmarks, bookmarksnumbered, colorlinks, linkcolor={red},citecolor={green}]{hyperref}
\begin{document}
\title{\bf Brane worlds and dark matter}
\author{Shahab Shahidi\footnote{Electronic address: s$\_$shahidi@mail.sbu.ac.ir}\quad and\quad Hamid Reza Sepangi\footnote{Electronic address:
hr-sepangi@sbu.ac.ir} \\
\small {Department of Physics, Shahid Beheshti University, G. C.,
Evin, Tehran (19839) Iran}}
\maketitle

\begin{abstract}
Two problems related to dark matter is considered in
the context of a brane world model in which the confinement of gauge
fields on the brane is achieved by invoking a confining potential.
First, we show that the virial mass discrepancy can be addressed if
the conserved geometrical term appearing in this model is considered
as an energy momentum tensor of an unknown type of matter, the
so-called X-matter whose equation of state is also obtained. Second,
the galaxy rotation curves are explained by assuming an anisotropic
energy momentum tensor for the X-matter.
\end{abstract}

\section{Introduction}
Over the past decade, model theories in which our $4D$ space-time
(brane) embedded as a hyper-surface in a higher dimensional
space-time (bulk) have been proposed and opened a large, relatively
unexplored area of research. The interest in the study of gravity in
spaces with extra dimensions started to grow rapidly with the
appearance of the proposal put forward by  Randall and Sundrum (RS)
\cite{rnd1,rnd2}. This was followed by the derivation of the
Einstein field equations on the brane through the use of the
Gauss-Codazzi equations and Israel junction conditions together with
the assumption of $\mathbb{Z}_2$ symmetry \cite{sms}. The
cosmological consequences of this work, such as the additional
quadratic density term in the Friedman equations \cite{brax,bine}
have been widely investigated.

Although use of the Israel junction conditions together with
$\mathbb{Z}_2$ symmetry is the widely accepted method amongst the
workers in the field for confining the gauge fields to the brane and
hence obtaining the field equation, there has been concerns
expressed over their use in that such junction conditions may not be
unique \cite{battye}. Another point to note is that if more than one
extra dimension is involved, there is, as yet, no way of obtaining
viable junction conditions. Eying such concerns, an interesting
model which avoids the above questions was proposed in \cite{maia1}.
Here, the confinement of the gauge fields to the brane is done by
postulating the existence of a confining potential $\mathcal{V}$.
One then obtains the field equations on the brane without the use of
any junction conditions or the assumption of $\mathbb{Z}_2$
symmetry. The field equations obtained in this way contain a term,
$Q_{\mu\nu}$, which is an independently conserved quantity and
geometrical in nature. This would allow us to attribute it to what
is known as the X-matter. The dynamics of test particles confined to
such a brane-world at the classical and quantum levels was studied
in \cite{jalal}. The static, spherically symmetric solutions of this
model was used subsequently to explain the galaxy rotation curves
\cite{heydar1}. Various other cosmological implication of the model
were also studied and showed to be consistent with observations,
such as the late time accelerated expansion of the universe which is
related to dark energy content of the universe.
\cite{heydar2,heydar3}.

One of the important observation which necessitates the existence of
dark matter is the so-called virial mass discrepancy. The total mass
of a cluster of galaxies can be estimated in two ways. Knowing the
motions of its member galaxies, the virial theorem gives one
estimate $M_{_V}$. The second estimate is obtained by taking the
mass of the individual galaxies separately and adding them up to
give $M$. Almost without any exception, the virial mass $M_{_V}$ is
greater than the total baryonic mass $M$ by 20-30\% \cite{binney},
so that the existence of dark matter is assumed in order to explain
this discrepancy. In \cite{har1}, this effect is explained within a
RS type brane-world together with the fact that the electric part of
the Weyl tensor can be decomposed irreducibly into dark matter
energy density and pressure, previously presented in \cite{maar}.
The virial mass discrepancy is explained in \cite{heydar4} for a
brane-world model theory discussed above, and also in the context of
DGP brane gravity \cite{Shahab}.

Recent observations show that the tangential velocity of stars
moving around the center of a galaxy tend to a constant value as we
move away from the center of the galaxy. In order to explain such
rotation curves various suggestions have been made. For example, in
\cite{beken,moffat} Modified Newtonian Dynamics is used to explain the
stability of circular orbits, while in \cite{har3,har4} the authors use
conformal symmetry in the context of brane-world gravity to explain
the constancy of galaxy rotation curves. In this paper we present
similar conclusions within the context of the brane-world scenario
proposed here by invoking the notion of conformal Killing symmetry.

Calculation of the geometrical term obtained in this model needs the
specification of the bulk geometry. One may then compute the tensor
$Q_{\mu\nu}$ using the Gauss-Codazzi equation. In the case that the
bulk geometry is unknown,  one of the components of the extrinsic
curvature tensor is left arbitrary, which may then be obtained using
the Einstein field equations \cite{maia2}. In this paper however, we
use this arbitrariness and consider the geometric tensor
$Q_{\mu\nu}$ as a perfect fluid for which the energy density and
pressure is obtained via the Einstein field equations. As is shown
in the sections below, the following two problems related to dark
matter can be explained  using different geometrical matter. The
galaxy rotation curves can be accounted for by defining the tensor
$Q_{\mu\nu}$ to be anisotropic with the equation of state of the
form (\ref{eqs}). The virial mass discrepancy can be explained by
introducing the isotropic geometrical matter. This choice however is
somewhat expected in that the virial mass problem concerns the
calculation of the mass of the cluster of galaxies which is a much
larger object than the galaxy itself.

The scope of the paper is as follows. In the next section we obtain
the virial mass theorem in the context of the brane-world model
discussed above. In section three the equation of state of the
X-matter is obtained with the specific choice of the energy momentum
tensor of the X-matter. Section four concerns the galaxy rotation
curves where we show that the constancy of the velocity  is obtained
by the assumption that the brane admits a conformal symmetry.
Conclusions are drawn in section five.

\section{Virial theorem}
We use the Einstein field equations obtained in \cite{heydar4} in
the context of brane-world models without the $\mathbb{Z}_2$
symmetry. Neglecting the global bulk effects, the Einstein equations
read
\begin{equation}
G_{\mu\nu}=8\pi G\tau_{\mu\nu}-\lambda\textrm{g}_{\mu\nu}+Q_{\mu\nu},\label{eq11}
\end{equation}
where $\tau_{\mu\nu}$ is the energy-momentum tensor of the ordinary
matter on the brane and $Q_{\mu\nu}$ is a conserved geometric
quantity which can be considered as an energy-momentum of the
X-matter. In this section we obtain the generalized virial theorem
and show that the X-matter mass is linearly related to the virial
mass. We take the brane spherically symmetric and static metric as
\begin{align}
ds^2=-e^{\mu(r)}dt^2+e^{\nu(r)}dr^2+r^2 (d\theta^2+\sin^2\theta)
d\varphi^2.\label{eq12}
\end{align}
For $\tau_{\mu\nu}$, we take
\begin{align}
\tau_{\mu\nu}=\mbox{diag}\left(-\rho_{eff},p^{(r)}_{eff},p^{(\bot)}_{eff},p^{(\bot)}_{eff}\right),
\label{eq13}
\end{align}
while for the X-matter we choose a fluid of the form
\begin{align}
Q_{\mu\nu}= 8\pi G \mbox{diag}\left(-\rho_{_X},p_{_X},p_{_X},p_{_X}\right).\label{eq14}
\end{align}
Now, substituting the equations above in equation (\ref{eq11}) we
obtain the following field equations on the brane
\begin{align}
&\frac{e^{-\nu(r)}}{r^2}(1-r\nu^\prime)-\frac{1}{r^2}=-8\pi G(\rho_{eff}+\rho_{_X})-\lambda,\label{eq15}\\
&\frac{e^{-\nu(r)}}{r^2}(1+r\mu^\prime)-\frac{1}{r^2}=8\pi G (p^{(r)}_{eff}+p_{_X})-\lambda,\label{eq16}\\
&\frac{e^{-\nu(r)}}{4r}(2\mu^\prime-2\nu^\prime-\mu^\prime\nu^\prime
r+2\mu^{\prime\prime}r+ \mu^{\prime ^2}r)=8\pi G
(p^{(\bot)}_{eff}+p_{_X})-\lambda,\label{eq17}\\
&\mu^\prime=-2\frac{\frac{d}{d
r}p^{(r)}_{eff}+p^\prime_{_X}+\frac{2}{r}
(p^{(r)}_{eff}-p^{(\bot)}_{eff})}{\rho_{eff}+p^{(r)}_{eff}+\rho_{_X}+p_{_X}},\label{eq18}
\end{align}
where the last equation is obtained from
$\nabla^\mu(Q_{\mu\nu}+ 8\pi G \tau_{\mu\nu})=0$ and a prime represents
derivative with respect to $r$.

In order to obtain the virial theorem, it is convenient to work in
tetrad formalism. Let us then define the following frame of
orthonormal vectors
\begin{align}
e^{(0)}_\rho=e^{\frac{\mu}{2}}\delta^0_\rho,\qquad
e^{(1)}_\rho=e^{\frac{\nu}{2}}\delta^1_\rho,\qquad
e^{(2)}_\rho=r\delta^2_\rho,\qquad e^{(3)}_\rho=r
\sin\theta\delta^3_\rho,\label{eq19}
\end{align}
where $\textrm{g}^{\mu\nu}e^{(a)}_\mu e^{(b)}_\nu=\eta^{(a)(b)}$,
and the tetrad indices are indicated by the parenthesis. In tetrad
components the 4-velocity $v^\mu$ of a typical galaxy, with the
property $v^\mu v_\mu=-1$, is given by
\begin{align}
v^{(a)}=v^\mu e^{(a)}_\mu,\qquad a=0,1,2,3.\label{eq20}
\end{align}
Finally we introduce $f(x^\mu,v^{(a)})$ as the distribution function
of galaxies assumed to be identical and collisionless point
particles. The relativistic Boltzmann equation in tetrad components
is now given by \cite{lind,jack}
\begin{align}
v^{(a)} e^\rho_{(a)}\frac{\partial f}{\partial
x^\rho}+\gamma^{(a)}_{(b)(c)}v^{(b)}v^{(c)}\frac{\partial
f}{\partial v^{(a)}}=0,\label{eq21}
\end{align}
where
$\gamma^{(a)}_{(b)(c)}=e^{(a)}_{\rho;\sigma}e^\rho_{(b)}e^\sigma_{(c)}$
are the Ricci rotation coefficients. For our  choice of the metric,
the Boltzmann equation becomes
\begin{align}
v_r\frac{\partial f}{\partial
r}&+e^{\frac{\nu}{2}}\frac{v_\theta}{r}\frac{\partial f}{\partial
\theta}+e^{\frac{\nu}{2}}\frac{v_\varphi}{r\sin\theta}\frac{\partial
f}{\partial \varphi}
-\left(\frac{v^2_t}{2}\mu^\prime-\frac{v^2_\theta+v^2_\varphi}{r}\right)\frac{\partial
f}{\partial v_r}-\frac{v_r}{r}\left(v_\theta \frac{\partial
f}{\partial v_\theta}+v_\varphi \frac{\partial f}{\partial
v_\varphi}\right) \nonumber \\
&-e^{\frac{\nu}{2}}\frac{v_\varphi}{r}\cot\theta\left(v_\theta
\frac{\partial f}{\partial v_\varphi}-v_\varphi \frac{\partial
f}{\partial v_\theta}\right)=0,\label{eq22}
\end{align}
where
\begin{align}
v^{(0)}=v_t,\qquad v^{(1)}=v_r,\qquad v^{(2)}=v_\theta,\qquad
v^{(3)}=v_\varphi.\label{eq23}
\end{align}
If we assume that the coordinate dependence of the distribution
function is only through $r$, equation (\ref{eq22}) reduces to
\begin{align}
v_r\frac{\partial f}{\partial
r}&-\left(\frac{v^2_t}{2}\mu^\prime-\frac{v^2_\theta+v^2_\varphi}{r}\right)\frac{\partial
f}{\partial v_r}-\frac{v_r}{r}\left(v_\theta \frac{\partial
f}{\partial v_\theta}+v_\varphi \frac{\partial f}{\partial
v_\varphi}\right) \nonumber \\
&-\frac{e^{\frac{\nu}{2}}v_\varphi}{r}\cot\theta\left(v_\theta
\frac{\partial f}{\partial v_\varphi}-v_\varphi \frac{\partial
f}{\partial v_\theta}\right)=0.\label{eq24}
\end{align}
The spherically symmetric nature of the metric requires that the
coefficient of $\cot\theta$ be zero. Now, multiplying equation
(\ref{eq24}) by $mv_r dv$, where $dv=\frac{1}{v_t}dv_r dv_\theta
dv_\varphi$ is an invariant volume element in the velocity space and
$m$ is the mass of the galaxy, and integrating over the velocity
space and assuming that $f$ vanishes sufficiently rapidly as the
velocities tend to $\pm\infty$, we obtain
\begin{align}
r\frac{\partial}{\partial
r}\left[\rho\left<v^2_r\right>\right]+\frac{1}{2}\rho\left[\left<v^2_t\right>+
\left<v^2_r\right>\right]r\mu^\prime
-\rho\left[\left<v^2_\theta\right>+\left<v^2_\varphi\right>-2\left<v^2_r\right>\right]=0,\label{eq25}
\end{align}
where  $\rho$ is the mass density, and $\left< \right>$ represents
the average value of the quantity it contains. Multiplying equation
(\ref{eq25}) by $4\pi r^2$ and integrating over the cluster of
galaxies, we obtain
\begin{align}
-\int_0^R
4\pi\rho\left[\left<v^2_r\right>+\left<v^2_\theta\right>+\left<v^2_\varphi\right>\right]r^2dr+
\frac{1}{2}\int_0^R 4\pi
r^3\rho\left[\left<v^2_t\right>+\left<v^2_r\right>\right]\frac{\partial\mu}{\partial
r}dr=0.\label{eq26}
\end{align}
The total kinetic energy of the galaxies is given by
\begin{align}
K=\int_0^R
2\pi\rho\left[\left<v^2_r\right>+\left<v^2_\theta\right>+\left<v^2_\varphi\right>\right]r^2dr,\label{eq27}
\end{align}
so that equation (\ref{eq26}) reduces to
\begin{align}
2K=\frac{1}{2}\int_0^R 4\pi
r^3\rho\left[\left<v^2_t\right>+\left<v^2_r\right>\right]\frac{\partial\mu}{\partial
r}d r. \label{eq28}
\end{align}
In terms of the distribution function we can write the energy
momentum tensor $\tau_{\mu\nu}$ as \cite{lind}
\begin{align}
\tau_{\mu\nu}=\int f m v_\mu v_\nu d v,\label{eq29}
\end{align}
leading to
\begin{align}
\rho_{eff}=\rho\left<v^2_t\right>,\qquad p^{(r)}_{eff}=\rho\left<v_r^2\right>,\qquad
p^{(\bot)}_{eff}=\rho\left<v_\theta^2\right>=\rho\left<v^2_\varphi\right>.
\label{eq30}
\end{align}
Now, by adding the minus of equation (\ref{eq15}) to equation
(\ref{eq16}) and adding twice of equation (\ref{eq17}) to the result we
obtain
\begin{align}
e^{-\nu}\left(\frac{\mu^\prime}{r}-\frac{\mu^\prime \nu^\prime}{4}+\frac{\mu^{\prime\prime}}{2}+\frac{\mu^{\prime ^2}}
{4}\right)=4\pi G\rho\left<v^2\right>+ 4\pi G  \left(\rho_{_X}+3p_{_X}\right)-\lambda, \label{eq31}
\end{align}
where we have defined
$\left<v^2\right>=\left<v_t^2\right>+\left<v_r^2\right>+\left<v_\theta^2\right>+
\left<v_\varphi^2\right>$. Let us now see what kind of
approximations we can make in this formalism. First, we assume that
$\mu(r)$ and $\nu(r)$ are small so that we can neglect the quadratic
terms in equation (\ref{eq31}). Second, we assume that the galaxies
have velocities much smaller than the speed of light, so
$\left<v_r^2\right>,\left<v_\theta^2\right>,
\left<v_\varphi^2\right> \ll \left<v_t^2\right> \approx 1$. These
conditions can be applied to test particles in stable circular
motion around galaxies and to the galactic clusters. Equation
(\ref{eq31}) is then reduced to
\begin{align}
4\pi G \rho=\frac 1 2 \frac{1}{r^2}\frac{\partial}{\partial r}\left(r^2\mu^\prime\right)+\lambda- 4\pi G
\left(\rho_{_X}+3p_{_X}\right).\label{eq32}
\end{align}
To go any further, we need to specify the equation of the state for
the X-matter which we assume to have the following form
\begin{align}
p_{_X}=\alpha\rho_{_X},\label{eq33}
\end{align}
where $\alpha$ is an arbitrary constant. Equation (\ref{eq32}) is
therefore reduced to
\begin{align}
4\pi G \rho=\frac 1 2 \frac{1}{r^2}\frac{\partial}{\partial r}\left(r^2\mu^\prime\right)+\lambda-
4\pi G (1+3\alpha)\rho_{_X}.\label{eq34}
\end{align}
Multiplying equation (\ref{eq34}) by $r^2$ and integrating from $0$
to $r$ we obtain
\begin{align}
\frac 1 2 r^2 \frac{\partial\mu}{\partial r}-GM(r)+\frac 1 3 \lambda r^3-GM_{_X}(r)=0,\label{eq35}
\end{align}
where we have defined
\begin{align}
M(r)=4\pi\int^r_0 \rho r^{\prime ^2} d r^\prime,\label{eq36}
\end{align}
and
\begin{align}
M_{_X}(r)=4\pi(1+3\alpha)\int^r_0 \rho_{_X} r^{\prime ^2} dr^\prime.\label{eq37}
\end{align}
Finally,  multiplying equation (\ref{eq35}) by $\frac{dM(r)}{r}$ and
integrating from $0$ to $R$ and using equation (\ref{eq28}) we
obtain
\begin{align}
W+2K+\frac 1 3 \lambda I+W_{_X}=0,\label{eq38}
\end{align}
where
\begin{align}
&W=-\int_0^R \frac{GM(r)}{r}dM(r),\label{eq39}\\
&W_{_X}=-\int_0^R \frac{GM_{_X}(r)}{r}dM(r),\label{eq40}
\end{align}
and
\begin{align}
I=\int^R_0 r^2 dM(r),\label{eq41}
\end{align}
is the moment of inertia of the system. Equation (\ref{eq38}) is the
generalized virial theorem on the brane in the presence of the brane
cosmological constant and X-matter. If $W_{_X}=0$, equation
(\ref{eq38}) reduces to the virial theorem with a cosmological
constant \cite{jack}. In what follows we assume that $\alpha\neq
-\frac 1 3$. In order to obtain the ratio of the X-matter mass to
the virial mass, we introduce the following radii \cite{jack}
\begin{align}
&R_{_V}=\frac{M^2}{\int^{_R}_{_0} \frac{M(r)}{r}dM(r)},\label{eq42}\\
&R^2_{_I}=\frac{\int^{_R}_{_0} r^2 dM(r)}{M(r)},\label{eq43}\\
&R_{_X}=\frac{M^2_{_X}}{\int^{_R}_{_0} \frac{M_{_X}(r)}{r} dM(r)},\label{eq44}
\end{align}
where $R_{_V}$ is the virial radius and $R_{_X}$ is the radius defined by the X-matter. If we define the virial mass
as
\begin{align}
2K=\frac{GM^2_{_V}}{R_{_V}},\label{eq45}
\end{align}
and use the following relations
\begin{align}
W=-\frac{GM^2}{R_{_V}},\qquad W_{_X}=-\frac{GM_{_X}^2}{R_{_X}},\qquad I=MR^2_{_I},\label{eq46}
\end{align}
the generalized virial theorem (\ref{eq38}) reduces to
\begin{align}
\left(\frac{M_{_V}}{M}\right)^2=1+\left(\frac{M_{_X}}{M}\right)^2 \left(\frac{R_{_V}}{R_{_X}}\right)
-\frac{\lambda}{3G}\frac{R_{_I}^2 R_{_V}}{M}.\label{eq47}
\end{align}
Since the contribution of a cosmological constant to the mass of the
galaxy is several order of magnitude smaller than the observed mass,
we can neglect it in equation (\ref{eq47}). Moreover we can
neglect the unity in (\ref{eq47}) since $M_{_V}$ is much greater
than $M$ for most galaxies. Therefore, the virial mass in our model
is given by
\begin{align}
M_{_V} \simeq M_{_X} \sqrt{\frac{R_{_V}}{R_{_X}}},\label{eq48}
\end{align}
showing that the virial mass is proportional to the X-matter mass.

\subsection{Estimating the virial mass}
In this section we obtain $M_{_X}$ as a function of $r$ and
consequently see that the solution is linearly increasing with $r$,
pointing to a possible explanation of the mass discrepancy in
clusters of galaxies. We start with Einstein equations
(\ref{eq15})-(\ref{eq18}) and since most of the baryonic mass in
clusters is in the gas form, we assume that the effective
energy-density and pressure in $\tau_{\mu\nu}$ is that of a gas and
therefore set
\begin{align}
\rho_{eff}=\rho_g(r),\qquad p^{(r)}_{eff}=p^{(\bot)}_{eff}=p_g(r).\label{eq49}
\end{align}

In majority of clusters most of the baryonic mass is in the form of
the intra-cluster gas. The gas density $\rho_g$ can be fitted with
the observational data by using the following radial baryonic mass
distribution \cite{har1,rei}
\begin{align}
\rho_g(r)=\rho_0\left(1+\frac{r^2}{r_c^2}\right)^{-\frac{3\beta}{2}},\label{eq63}
\end{align}
where $r_c$ is the core radius, and $\rho_0$ and $\beta$ are cluster
independent constants. A static spherically symmetric system of
collisionless particles that is in equilibrium, can be described by
the Jean's equation \cite{binney}
\begin{align}
\frac{d}{dr}\left[\rho_g\sigma^2_r\right]+\frac{2\rho_g(r)}{r}\left(\sigma^2_r-
\sigma^2_{\theta,\varphi}\right)=
-\rho_g(r)\frac{d\Phi}{dr},\label{eq64}
\end{align}
where $\Phi(r)$ is the gravitational potential, and $\sigma_r$ and
$\sigma_{\theta,\varphi}$ are the mass-weighted velocity
dispersions, respectively, in the radial and tangential directions.
We assume that the gas is isotropically distributed inside the
cluster, so $\sigma_r=\sigma_{\theta,\varphi}$. The gas pressure is
related to the velocity dispersion and gas density by
$p_g=\rho_g\sigma^2_r$. By assumption that the gravitational field
is weak so that it satisfies the usual Poisson equation
$\Delta\Phi\approx 4\pi\rho_{tot}$, where $\rho_{tot}$ is the energy
density including $\rho_g$ and other forms of matter, like luminous
matter and the X-matter, etc., the Jean's equation becomes
\begin{align}
\frac{dp_g(r)}{dr}=-\rho_g(r)\frac{d\Phi}{dr}=-\frac{GM_{tot}}{r^2}\rho_g(r),\label{eq65}
\end{align}
where $M_{tot}(r)$ is the total mass inside the radius $r$. The
observed x-ray emission from the hot ionized intra-cluster gas is
usually interpreted by assuming that the gas is in isothermal
equilibrium. So we assume that the gas is in equilibrium state
having the equation of state \cite{jack}
\begin{align}
p_g(r)=\frac{k_{_B}T_g}{\mu m_p}\rho_g(r),\label{eq66}
\end{align}
where $\mu=0.61$ is the mean atomic weight of the particles in the
gas cluster, and $m_p$ is the proton's mass. Equation (\ref{eq65})
then becomes
\begin{align}
M_{tot}(r)=-\frac{k_{_B}T_g}{\mu m_p G}~r^2\frac{d}{dr}\ln\rho_g=
\frac{3k_{_B}T_g\beta}{\mu m_p G}\frac{r^3}{r^2+r_c^2},
\label{eq67}
\end{align}
where the second equality is obtained using equation (\ref{eq63}).
Using equations (\ref{eq36}) and (\ref{eq37}) we can obtain another
expression for $M_{tot}$
\begin{align}
\frac{dM_{tot}}{dr}=4\pi r^2\rho_g+4\pi (1+3\alpha)r^2\rho_{_X}.\label{eq68}
\end{align}
Combining equations (\ref{eq67}) and (\ref{eq68}) we obtain the following expression for $\rho_{_X}$
\begin{align}
\rho_{_X}=\frac{1}{1+3\alpha}\left[\frac{3k_{_B}T_g\beta}{4\pi\mu m_p G}~\frac{r^2+3r^2_c}{(r^2+r^2_c)^2}-
\rho_0\left(1+\frac{r^2}{r^2_c}\right)^{-\frac{3\beta}{2}}\right].\label{eq69}
\end{align}
In the regime $r \gg r_c$, equation (\ref{eq69}) is reduced to
\begin{align}
\rho_{_X}\simeq\frac{1}{1+3\alpha} \left[\frac{3k_{_B}T_g\beta}{4\pi\mu m_p G}-
\rho_0 r_c^{3\beta} r^{2-3\beta} \right]\frac{1}{r^2},\label{eq70}
\end{align}
and by equation (\ref{eq37}) we obtain
\begin{align}
M_{_X} \simeq \left[\frac{3k_{_B}T_g\beta}{\mu m_p G}-\frac{4\pi\rho_0 r_c^{3\beta}}{3(1-\beta)}
r^{2-3\beta} \right]r.\label{eq71}
\end{align}
We now consider the limit of large $r$. For most clusters the
parameter $\beta$ has the value $\beta\geq\frac 2 3$ \cite{rei}.
Because of the smallness of the second term in (\ref{eq71}) in this
case, we have
\begin{align}
M_{_X}\approx \frac{3k_{_B}\beta T_g}{\mu m_p G}~r.\label{eq711}
\end{align}
As can be seen, the X-matter mass increases linearly with $r$ and
this is similar to the behavior of dark matter in the cluster of
galaxies. Let us now make some estimates. First we note that
$k_{_B}T_g\approx 5~KeV$ for most clusters. The virial radius of
clusters is usually assumed to be $r_{200}$, indicating the radius
for which the energy density $\rho_b$ of the cluster becomes
$\rho_{200}=200\rho_{univ}$, where $\rho_{univ}=4.6975\times
10^{-30} h_{50}^2~g/cm^3$ \cite{rei}. We can estimate the critical
radius of the X-matter by determining the distance in which the
X-matter density is equal to $\rho_{200}$. Using (\ref{eq70}) and
noting that the last term may be dropped in our limit we find
\begin{align}
r_{cr}=1.82\sqrt{\beta}\left(\frac{k_{_B}T_g}{5~KeV}\right)^{\frac{1}{2}}
h_{50}^{-1}~Mpc. \label{eq71.1}
\end{align}
We then have
\begin{align}
M^{cr}_{_X}=9.8\times 10^{14}~\beta^{\frac 3 2}\left(\frac{k_{_B}T_g}
{5~KeV}\right)^{\frac 3 2}h_{50}^{-1}M_\odot. \label{eq72}
\end{align}
 For $\beta<\frac 2 3$ the X-matter mass has a maximum at
\begin{align}
R_{max}=\left(\frac{3 k_{_B}\beta T_g}{4\pi G\rho_0\mu m_p r_c^{3\beta}}
\right)^\frac{1}{2-3\beta}.\label{eq74}
\end{align}
The maximum value of the X-matter mass is given by
\begin{align}
GM_{_X}=\left(\frac{3 k_{_B}\beta T_g}{\mu m_p G}\right)\left(\frac{2-3\beta}
{3(1-\beta)}\right)~R_{max}.\label{eq75}
\end{align}

\section{X-matter solutions on the brane}
In this section we find the vacuum solutions of the theory and
obtain the equation of motion of the X-matter. Assuming that
$\tau_{\mu \nu}=0$, the effective vacuum equation (\ref{eq11}) is
reduced to
\begin{equation}
G_{\mu\nu}=-\lambda \textrm{g}_{\mu\nu} +Q_{\mu\nu}. \label{1eq23}
\end{equation}
Let us assume that $Q_{\mu\nu}$ has the form of an anisotropic perfect fluid
\begin{equation}
Q^\mu_{~\nu}=8\pi G\,\,
\mbox{diag}\left(-\rho_{_X},p^{\parallel}_{_X},p^{\perp}_{_X},p^{\perp}_{_X}\right),
\label{1eq24}
\end{equation}
with an equation of state
\begin{equation}
\rho_{_X}=p^{\parallel}_{_X}. \label{eqs}
\end{equation}
Taking  metric (\ref{eq12}), the vacuum field equations on the brane become
\begin{align}
&\frac{e^{-\nu(r)}}{r^2}(1-r\nu^\prime)-\frac{1}{r^2}=-\lambda-8\pi G \rho_{_X}\label{1eq29}\\
&\frac{e^{-\nu(r)}}{r^2}(1+r\mu^\prime)-\frac{1}{r^2}=-\lambda+8\pi G p^{\parallel}_{_X},\label{1eq30}\\
&\frac{e^{-\nu(r)}}{4r}(2\mu^\prime-2\nu^\prime-\mu^\prime\nu^\prime
r+2\mu^{\prime\prime}r+ \mu^{\prime\, 2}r)=-\lambda+8\pi G
p^{\perp}_{_X}.\label{1eq31}
\end{align}
The equations above contain 4 unknown quantities so that an extra
relation would be desirable. Such an equation can be provided by
assuming that the brane admits a one parameter group of conformal
motions so that
\begin{equation}
\mathcal{L}_{\xi}\textrm{g}_{\mu\nu}=\phi(r)\textrm{g}_{\mu\nu},
\label{1eq32}
\end{equation}
where $\phi(r)$ is the conformal factor and $\mathcal{L}$ represents
the Lie derivative. Moreover, we assume the following general form
for the vector field $\xi$
\begin{equation}
\xi =\xi^{0}(t,r)\frac{\partial}{\partial t}+\xi^{1}(t,r)\frac{\partial}{\partial r}+
\xi^{2}(\theta,\varphi)\frac{\partial}{\partial\theta}+\xi^{3}(\theta,\varphi)\frac{\partial}{\partial\varphi}.
\label{1eq33}
\end{equation}
Use of equation  (\ref{1eq32}) then leads to
\begin{equation}
\mu^\prime \xi^1+2\frac{\partial\xi^0}{\partial t}=\phi(r),\label{1eq34}
\end{equation}
\begin{equation}
\nu^\prime \xi^1+2\frac{\partial\xi^1}{\partial r}=\phi(r),\label{1eq35}
\end{equation}
\begin{equation}
e^\nu\frac{\partial\xi^1}{\partial t}=e^\mu\frac{\partial\xi^0}{\partial r},\label{1eq36}
\end{equation}
\begin{equation}
\frac{1}{r}\xi^1+\frac{\partial\xi^2}{\partial\theta}=\frac{\phi(r)}{2},\label{1eq37}
\end{equation}
\begin{equation}
\frac{1}{r}\xi^1+\cot\theta\xi^2+\frac{\partial\xi^3}{\partial\varphi}=\frac{\phi(r)}{2},\label{1eq38}
\end{equation}
\begin{equation}
\sin^2\theta\frac{\partial\xi^3}{\partial\theta}=-\frac{\partial\xi^2}{\partial\varphi}.\label{1eq39}
\end{equation}
In order to solve these equations, we see from equation  (\ref{1eq37}) that
\begin{align}
&\xi^2=\frac{dF(\varphi)}{d\varphi},\label{1eq40}\\
&\xi^1=\frac{r}{2}\phi,\label{1eq41}
\end{align}
where we have set the arbitrary separation constant equal to zero
(this can be done easily by shifting $\phi$ by a constant) and
$F(\varphi)$ is an arbitrary function. From equation  (\ref{1eq36})
we have $\xi^0=\xi^0(t)$ which, upon using equation (\ref{1eq34}), we
obtain
\begin{equation}
\xi^0=\frac{k}{2} t+A,\label{1eq42}
\end{equation}
where $k$ is an arbitrary separation constant and $A$ is some other
constant. Now, using equation  (\ref{1eq38}) we obtain
\begin{equation}
\xi^3=-\cot\theta F(\varphi)+G(\theta),\label{1eq43}
\end{equation}
where $G(\theta)$ is some arbitrary function. Since
$\frac{\partial}{\partial t}$ is a killing vector field we can set
$A=0$ without loss of generality, so that the form of the conformal
killing vector field $\xi$ is given by
\begin{align}
\xi &=\frac{k}{2}t\frac{\partial}{\partial
t}+\frac{r\phi}{2}\frac{\partial}{\partial r}+
\frac{dF(\varphi)}{d\varphi}\frac{\partial}{\partial\theta}
-\left[\cot\theta
F(\varphi)-G(\theta)\right]\frac{\partial}{\partial\varphi}.\label{1eq44}
\end{align}
Now, from equations (\ref{1eq34}) and (\ref{1eq35}) we can write the
metric components in terms of the conformal factor
\begin{align}
&e^{\mu(r)}=C^2 r^2 exp \left[-2k\int\frac{dr}{r\phi}\right],\label{1eq45}\\
&e^{\nu(r)}=\frac{B^2}{\phi^2},\label{1eq46}
\end{align}
where $C$ and $B$ are arbitrary integration constants. Upon
substitution of these equations into the field equations (\ref{1eq29})-(\ref{1eq31})
we obtain the following system of equations for $\phi(r)$ , $\rho_{_X}$
and $p^{\perp}_{_X}$
\begin{align}
&\frac{1}{r^2}\frac{\phi^2}{B^2}\left(1+2r\frac{\phi^\prime}{\phi}\right)-\frac{1}{r^2}=
-\lambda-8\pi G \rho_{_X},\label{1eq47}\\
&\frac{1}{r^2}\frac{\phi^2}{B^2}\left(3-2\frac k \phi \right)-\frac{1}{r^2}=
-\lambda+8\pi G\rho_{_X},\label{1eq48}\\
&\frac{1}{r^2}\frac{\phi^2}{B^2}\left(2r\frac{\phi^\prime}{\phi}+\frac{(\phi-k)^2}{\phi^2}\right)=
-\lambda+8\pi G p^{\perp}_{_X}.
\label{1eq49}
\end{align}
Assuming $\lambda=0$ and equating equations (\ref{1eq47}) and (\ref{1eq48}) we obtain
\begin{align}
r\phi\phi^\prime+2\phi^2-k\phi-B^2=0.\label{1eq50}
\end{align}
Now, if one expresses $r$ in terms of $\phi$, one finds
\begin{align}
r^2=r_0^2\frac{f(\phi)}{\sqrt{2\phi^2-k\phi-B^2}},\label{sol1}
\end{align}
where $r_0$ is an integration constant and $f(\phi)$ is defined as
\begin{align}
f(\phi)=\exp\left\{\frac{k}{\sqrt{8B^2+k^2}}\tanh^{-1}\left[\frac{-k+4\phi}
{\sqrt{8B^2+k^2}}\right]\right\}.
\label{sol2}
\end{align}
The energy density and pressure of the X-matter can now be obtained
in terms of $\phi$
\begin{align}
&8\pi G \rho_{_X}=8\pi G p^{\parallel}_{_X}=-\frac{1}{B^2 r_0^2}\frac{\sqrt{2\phi^2-k\phi-B^2}}{f(\phi)}
\left(B^2-3\phi^2+2k\phi\right) ,\label{1eq54}
\end{align}
and
\begin{align}
&8\pi G p^{\perp}_{_X}=-\frac{1}{B^2 r_0^2}\frac{\sqrt{2\phi^2-k\phi-B^2}}{f(\phi)}
\left(3\phi^2-2B^2-k^2\right).\label{1eq55}
\end{align}
The equation of state of the X-matter can
be obtained from equations (\ref{1eq54}) and (\ref{1eq55})
\begin{equation}
8\pi G(p^{\perp}_{_X}+\rho_{_X})=-\frac{1}{B^2 r^2}\left(k^2+2k\phi+B^2\right). \label{1eq57}
\end{equation}
We see that in the limit $r\rightarrow\infty$ the X-matter has an
equation of state $p^{\perp}_{_X}=-\rho_{_X}$.

\section{Galaxy rotation curves}
The results of the previous section is in agreement with the
observed behavior of the tangential velocity of galaxies; we know
from observations that the rotational velocities increase almost
linearly from the center of a galaxy and approaches a constant value
of about $200km/s$ as one moves away from the center \cite{binney}.
In this section we consider the tangential velocity of a test
particle which moves in a circular time-like geodesic orbit. The
Lagrangian of the system is given by \cite{mato}
\begin{eqnarray}
2L =\left(\frac{d s}{d\tau}\right)^2 =-e^{\mu(r)}\left(\frac{d
t}{d\tau}\right)^2+ e^{\nu(r)}\left(\frac{d
r}{d\tau}\right)^2+r^2\left(\frac{d\Omega}{d\tau}\right)^2,
\label{1eq58}
\end{eqnarray}
where $\tau$ is the affine parameter along the geodesic. From
equation (\ref{1eq58}) we find that the energy $E=e^\mu \dot{t}$, the
$\varphi$ component of the angular momentum of the particle
$l_\varphi=r^2\sin^2\theta\dot{\varphi}$, where a dot denotes
differentiation with respect to $\tau$ and the total angular
momentum of the particle $l^2=l_\theta^2+(l_\varphi / \sin\theta)^2$
are conserved quantities. The total angular momentum of the particle
can be written in terms of the solid angle as
$l^2=r^4\dot{\Omega}^2$ \cite{mato}. The equation of the geodesic
orbits can then be written as
\begin{align}
\dot{r}^2+V(r)=0,\label{1eq59}
\end{align}
where
\begin{align}
V(r)=-e^{-\nu}\left(E^2 e^{-\mu}-\frac{l^2}{r^2}-1\right).\label{1eq60}
\end{align}
For stable circular orbits, we must have
\begin{align}
\dot{r}=0,\qquad \frac{\partial V}{\partial r}=0,\qquad \frac{\partial^2 V}{\partial r^2}>0, \label{1eq61}
\end{align}
so that the potential describes a minimum of the motion. These
conditions lead to the following expressions for the energy and
total angular momentum as \cite{har3}
\begin{align}
E^2=\frac{2e^\mu}{2-r\mu^\prime},\qquad
l^2=\frac{r^3\mu^\prime}{2-r\mu^\prime}. \label{1eq62}
\end{align}
On the other hand the line element (\ref{eq12}) can be written in
terms of the velocity, measured by an inertial observer far from the
source as \cite{mato}
\begin{align}
ds^2=-dt^2\left(1-v^2\right), \label{1eq63}
\end{align}
where
\begin{align}
v^2=e^{-\mu}\left[ e^\nu \left(\frac{dr}{dt}\right)^2+r^2\left(\frac{d\Omega}{dt}\right)^2\right].
\label{1eq64}
\end{align}
\begin{figure}
  \centering
  \includegraphics{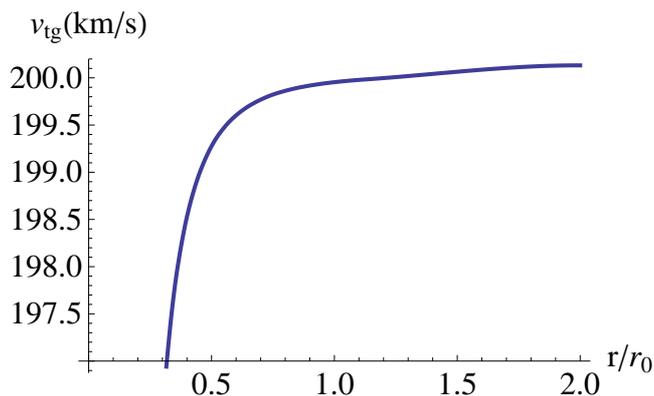}
\caption{\footnotesize Variation of $v_{tg}$ as a function of $r/r_0$ with $B=1.000000667$, and $k=1$.}
  \label{fig}
\end{figure}
In the case of stable circular orbits we have the following
expression for the tangential velocity \cite{lake}
\begin{align}
v^2_{tg}=r^2 e^{-\mu} \left(\frac{d\Omega}{dt}\right)^2.
\label{1eq65}
\end{align}
Using the expressions for the conserved quantities $E$ and $l^2$, we
obtain
\begin{align}
v^2_{tg}=\frac{e^\mu}{r^2}\frac{l^2}{E^2}.
\label{1eq66}
\end{align}
Use of equation (\ref{1eq62}) then leads to
\begin{equation}
v^2_{tg}=\frac{r\mu^\prime}{2},
\label{1eq67}
\end{equation}
showing that $v_{tg}$ depends only on the time-time component of the
metric. Using equation (\ref{1eq45}), the tangential velocity in terms of
$\phi$ is given by
\begin{equation}
v^2_{tg}=1-\frac{k}{\phi}.\label{1eq68}
\end{equation}
Although we cannot express $v_{tg}$ explicitly in terms of $r$,  we
can obtain the value of the tangential velocity at infinity from
equation (\ref{sol1}). We see from this equation that in the case
of $\phi\rightarrow\phi_{1,2}$ where
\begin{align}
\phi_{1,2}=\frac{k\pm\sqrt{k^2+8B^2}}{4}, \label{phi12}
\end{align}
we have $r\rightarrow\infty$. If we use the upper sign the
tangential velocity at infinity is given by
\begin{align}
v_{tg\infty}=\sqrt{1-\frac{4k}{k+\sqrt{8B^2+k^2}}}.\label{vinf}
\end{align}
We only need to fix the ratio of the constants in equation (\ref{vinf})
to be consistent with the observational data. For example with the
choice $B/k=1.000000667$ we have $v_{tg\infty}=200km/s$. Figure
\ref{fig} shows a plot of $v_{tg}$ as a function of $r$.

Let us obtain the X-matter mass of the galaxy and its dependence on
$r$. After integrating equation (\ref{1eq29}), we obtain
\begin{align}
e^{-\nu}=1-\frac{2G}{r}M_{_X}(r),\label{1eq70}
\end{align}
where we have defined
\begin{align}
M_{_X}(r)=4\pi\int^r_0 \rho_{_X}(r^\prime)r^{\prime 2}dr^\prime.
\label{1eq71}
\end{align}
That is equation (\ref{eq37}) with $\alpha=0$. Now define a boundary
$R$ of the galaxy where $\rho_b\approx 0$. We then have
\begin{align}
e^\mu=e^{-\nu}=1-\frac{2GM_b}{R},\label{1eq74}
\end{align}
where
\begin{align}
M_b=4\pi\int^R_0\rho(r^\prime)r^{\prime 2}dr^\prime ,
\label{1eq75}
\end{align}
is the total mass of the the galaxy. By equations (\ref{1eq46}),
(\ref{1eq68}) and (\ref{1eq74}), we can find the relation between
constants of the model in terms of the tangential velocity at the
boundary
\begin{align}
\frac{B^2}{k^2}=\frac{1-\frac{2GM_b}{R}}{[1-v^2_{tg}(R)]^2}.\label{1eq76}
\end{align}
Now, substituting equations (\ref{1eq46}) and (\ref{1eq68}) into equation (\ref{1eq70}) and using equation (\ref{1eq76}), we obtain
\begin{align}
M_{_X}=\frac{M_b}{R}~r,\label{1eq80}
\end{align}
where we have used the fact that the tangential velocity is much
smaller than the speed of light. The X-matter mass depends linearly
on $r$ even after the boundary of the galaxy is reached, and has a
value of the order of the baryonic mass. This is in agreement with
the observational data.

\section{Conclusions}
In this paper we have considered the problems of virial mass and
rotation curves of galaxies in the context of a brane-world scenario
which uses a confining potential instead of $\mathbb{Z}_2$ symmetry
and the Israel junction conditions in order to confine the gauge
fields to the brane. We have shown that these problems can be
adequately addressed in this model by identifying the conserved
geometric quantity $Q_{\mu\nu}$ with a new kind of matter, also
known as the X-matter. The virial theorem was then obtained,
assuming a perfect fluid form for $Q_{\mu\nu}$ and shown to  be
related linearly to a geometrical mass obtained by $Q_{\mu\nu}$.  We
also assumed an anisotropic form for $Q_{\mu\nu}$ in order to find
the energy-density and the pressure of  the X-matter and obtained
the tangential velocity of a point particle in the galaxy, showing
that it behaves linearly with respect to $r$. The X-matter mass of
the galaxy was then calculated and shown to be of the order of the
baryonic mass of the galaxy which extends beyond its boundary linearly
with $r$.


\end{document}